\documentclass[twocolumn,showpacs,preprintnumbers]{revtex4}
\usepackage{amssymb}
\usepackage{amsfonts}
\usepackage{amsmath}
\usepackage{graphicx}
\usepackage{dcolumn}
\usepackage{bm}

\input{tcilatex}

\begin{document}

\title{The \ Discrete-Continuous Logic and its possible quantum realizations}
\author{E. D. Vol}
\email{vol@ilt.kharkov.ua}
\affiliation{B. Verkin Institute for Low Temperature Physics and Engineering of the
National Academy of Sciences of Ukraine 47, Lenin Ave., Kharkov 61103,
Ukraine.}
\date{\today }

\begin{abstract}
We propose a new version of generalized probabilistic propositional logic,
namely, discrete-continuous logic (DCL) in which every generalized
proposition (GP) is represented as $2\times 2$ nondiagonal positive matrix
with unit trace. We demonstrate that on the set of propositions of this kind
one can define both the discrete logical operations (connectives) such as
negation and strong logical disjunction and in addition one parameter group
of continuous operations (logical rotations). We prove that an arbitrary
classical proposition (which in this logic is represented by the purely
diagonal matrix) can be considered as the result of strong disjunction of
two identical GP. This fact gives one a good reason to presume the DCL as a
prime logical substructure underlying to ordinary propositional logic, which
is recorded by our consciousness. We believe that proposed version of DCL
will find many applications both in physics (quantum logic) and also in
cognitive sciences (mental imagery) for better understanding of the pecular
nature of mental brain operations.
\end{abstract}

\maketitle

\section{Introduction}

The Boolean propositional logic is the very \ structure that on the one hand
accompanies to a certain extent all behavioral acts of most people in
everyday life and on the other hand can be considered as the necessary
framework for various scientific theories including the most subtle ones.
Turning to physics we note that the set of all two-valued quantities
describing some classical system (where these quantities admit joint
measurement) can be arranged exactly in the form of the Boolean algebra of
corresponding observables.It concerns not only deterministic observables but
probabilistic two- valued variables ( that can take two possible values with
certain probabilities) as well.In the latter case one must certainly use the
probabilistic generalization of Boolean algebra. However, since the
pioneering paper of Birkhoff and von Neumann \cite{1s} it is known that in
quantum mechanics the set of noncommuting observables can not be described
in simple terms of Boolean algebra and some generalization of it, that is
"quantum logic", is necessary. Although many ingenious attempts have been
made both to formulate the universal version of quantum logic, that , by
natural way, would solve famous quantum paradoxes of the Schrodinger cat
type and in addition allows one to consider consistently the problem of
measurement (see e.g. \cite{2s} and references therein). Unfortunately , to
the best of my knowledge, similar program to the bitter end was not realized
by anyone. Besides, there is another important reason \ for the
generalization of the Boolean logic (ordinary or probabilistic) arising from
famous pecularities of human brain activity. As it is well known (see e.g. 
\cite{3s}) the brain consists of two hemispheres which are roughly of equal
size and surface and at first glance are similar to each other. However,
unlike from other pair human organs these two hemispheres do not represent
information in an identical manner but perform specific cognitive and mental
functions. It is possible to take for granted that the left hemisphere (LH)
produces sequental processing of information in discrete units. In
particular it is responsible for the implementation of various logical
operations and for speech linguistic capacities as well. On the other hand
the right hemisphere (RH) produces the simultaneous and coherent processing
of information and it is responsible for such brain functions as for example
image formation, space imagination and for the music perception as well .
Summing up one can say: the Boolean logic par excellence is the logic only
of the LH. Hence natural question arises: whether there is the
generalization of Boolean logic that takes into account the possibility of
execution both discrete and continuous logical operations. In present paper
we propose the version of DCL that combines both these two types of
operations. This logic, in our opinion, could essentially expand the
possibilities of currently existing logical devices. It must be emphasized
however that in this paper we do not pretend to give the consistent
description of real brain activity. Rather our paper should be considered as
an attempt to offer the simple phenomenological model which would reflect
certain specific pecularities of brain logic. On the other hand we insist
that our approach differs from the most of standard constructions of
artificial intellect which as a rule are based on the Boolean logic only.
The further part of the paper is organized as follows. In chapter 2 we
briefly recall some results of our paper \cite{4s} which are necessary for
understanding \ of the present text. The main idea of our approach to logic
consists in the possibility to represent all plausible propositions of
Boolean probabilistic logic by $2\times 2$ diagonal positive matrices with
unity trace. These representative matrices of propositions can be naturally
considered as corresponding density matrices of relevant two level quantum
systems. Moreover, it turns out that all logical operations with plausible
propositions(that is logical connectives) can be obtained with the help of
positive definite transformations of the type, that we specify exactly. In
the chapter 3 which is the basic part of the present paper, we generalize
all constructions mentioned above on the case of propositions of more
general type (GP) that are now represented by nondiagonal positive $2\times
2 $ matrices of special form with unit trace . It turns out that in this
case apart from few discrete connectives there is also one parameter group
of continuous transformations in the space of all GP. For this reason we
presume that the " power" of DCL is much stronger compared with the "power"
of the ordinary Boolean logic and, in our opinion ,bringing it \ closer to
the rules of real brain logic. Finally in chapter 4 of the paper we
demonstrate that all concepts of DCL may be properly implemented in relevant
quantum systems by the methods of modern quantum engineering. We believe
therefore that the manifesto of R. Landauer "Information is physical" can be
applied to logic as well as to information.

Now let us begin to the concrete presentation of results of the paper.

\section{Preliminary information}

Let us briefly remind the necessary results of our paper \cite{4s} in which
the main concepts of probabilistic logic were formulated based on the point
of view of quantum theory of open systems (QTOS). In \cite{4s} we considered
the set of classical plausible propositions (CP), that is, the propositions
a truth or falsity of which are known not exactly, but only with certain
probability. It is possible to represent plausible propositions with the
help of diagonal $2\times 2$ matrices with positive elements the sum of
which is equal to unity. So, every plausible proposition $A$ can be
represented by its matrix $\rho \left( A\right) =%
\begin{pmatrix}
p_{A} & 0 \\ 
0 & 1-p_{A}%
\end{pmatrix}%
$ , where \ $p$ is the probability for $A$ to be true. In what follows, if
there does not lead to confusion, we will sometimes identify plausible
propositions with their representative matrices. It turns out that all
logical connectives between propositions in this matrix language can be
expressed by unified way (that has its roots in QTOS) by some positive
definite transformations of representative matrices conserving their
diagonal form and traces. Referring the reader for details to \cite{4s} we
present here only the essence of the approach applied, using some simple
examples. So, in order to write down the representative matrix for negation
of some proposition $A$, that is $\overline{A}$, one must perform the next
transformation: $\rho (\overline{A})=G_{N}\cdot \rho \left( A\right) \cdot
G_{N}^{T}=%
\begin{pmatrix}
1-p_{A} & 0 \\ 
0 & p_{A}%
\end{pmatrix}%
$, where $G_{N}=%
\begin{pmatrix}
0 & 1 \\ 
1 & 0%
\end{pmatrix}%
$ (as usually we denote as $A^{T}$ a matrix transposed to $A$). In a like
manner the representative matrix of arbitrary two- place logical connective
can be written in the form: $\rho _{C}=G_{C}\left[ \rho \left( A\right)
\otimes \rho \left( B\right) \right] G_{C}^{T}$, where$\rho \left( A\right)
\otimes \rho \left( B\right) $ is the tensor product of matrices
corresponding to propositions $A$ and $B$, and $G_{C}$ is a certain $2\times
4$ matrix which posseses two defining properties: 1) every element of $G_{C}$
is equal $1$ or $0$ and 2) in each column of $\ G_{C}$ only single element
is equal to$1$ and all the rest are equal to zero. One can verify directly
that these two properties of $G_{C}$ exactly ensure the positivity and
diagonality of $\rho _{C}$ and also the conservation of its trace. It is
clear also that with the help of specified transformations (we call them
admissible transformations) one can obtain all logical connectives between
any number of propositions as well. Thus any statements relating to
probabilistic Boolean propositions can be translated into the language of
admissible transformations with the density matrices of relevant two level
quantum systems. This language also let one to generalize approach proposed
on the case of more extensive class of logical propositions and possible
operations with them. Now let us pass to this main topic of the present
paper.

\section{Generalized propositions and logical operations with them}

First of all we want to specify in what sense the concept of classical
probabilistic propositions can be expended on more general case. With this
end in view let us consider generic nondiagonal $2\times 2$ positive matrix
with unit trace - $\rho \left( A\right) =%
\begin{pmatrix}
p_{A} & z_{A} \\ 
z_{A}^{\ast } & 1-p_{A}%
\end{pmatrix}%
$ and try to interpret it as a representative matrix of certain logical
proposition. To realize this idea it is necessary to define basic logical
connectives on the set of these matrices. Obviously that there is no problem
to define negation of proposition $\ A$. We can perfom this task by the same
transformation $G_{N}=%
\begin{pmatrix}
0 & 1 \\ 
1 & 0%
\end{pmatrix}%
$ as in previous chapter that leads to the required result:%
\begin{equation}
\overline{A}=G_{N}\cdot A\cdot G_{N}^{T}=%
\begin{pmatrix}
1-p_{A} & z_{A}^{\ast } \\ 
z_{A}^{{}} & p_{A}%
\end{pmatrix}%
.  \label{b1}
\end{equation}%
Unfortunately when we try to use the similar method for the definition of
arbitrary two place connectives we are faced with insuperable difficulties
since the transformations that were admissible for diagonal matrices in
nondiagonal case do not conserve the traces of corresponding tensor
products. So, we are needed to use another way for the decision of this
problem. To this end let us include in our consideration as representative
only the matrices of the special form, namely:%
\begin{equation}
\rho (\ A)=%
\begin{pmatrix}
p_{A} & i\alpha _{A} \\ 
-i\alpha _{A} & 1-p_{A}%
\end{pmatrix}%
,  \label{b2}
\end{equation}%
where $\alpha _{A}$ is some real number. Now one can define correctly the
next two-place connective between two such matrices $A=%
\begin{pmatrix}
p & i\alpha \\ 
-i\alpha & 1-p%
\end{pmatrix}%
$ and $B=%
\begin{pmatrix}
q & i\beta \\ 
-i\beta & 1-\beta%
\end{pmatrix}%
$ using the positive definite $2\times 4$ transformation $G_{\Delta }=%
\begin{pmatrix}
0 & 1 & 1 & 0 \\ 
1 & 0 & 0 & 1%
\end{pmatrix}%
$. As a result we obtain the proposition $(A\Delta B)$ with the
representative matrix:%
\begin{equation}
\rho (A\Delta B)=G_{\Delta }\left( A\otimes B\right) G_{\Delta }^{T}.
\label{b3}
\end{equation}%
It is easy to check directly that representative matrix $\rho (A\Delta B)$
has the next explicit form:%
\begin{widetext}
\begin{equation}
\rho \left( A\Delta B\right) =%
\begin{pmatrix}
p+q-2pq+2\alpha \beta  &  & i\alpha \left( 1-2q\right) +i\beta \left(
1-2p\right)  &  \\ 
-i\alpha \left( 1-2q\right) -i\beta \left( 1-2p\right)  &  & 
1-p-q+2pq-2\alpha \beta  & 
\end{pmatrix}%
.  \label{b4m}
\end{equation}
\end{widetext}Thus one can see that the matrix $\rho \left( A\Delta B\right) 
$ has the required form Eq. (\ref{b2})\ and hence belongs to the set of
generalized propositions. In the special case when $\alpha =\beta =0$
propositions $A$, $B$ and $A\Delta B$ become ordinary plausible ones. This
means that their representative matrices take diagonal form. Corresponding
classical plausible proposition $(A_{c}\Delta B_{c})$ reduces to the strong
disjunction \ of two probabilistic propositions, namely:%
\begin{equation}
\rho (A_{c}\Delta B_{c})=%
\begin{pmatrix}
p+q-2pq &  & 0 &  \\ 
0 &  & 1-p-q+2pq & 
\end{pmatrix}%
.  \label{b5}
\end{equation}%
(Remind here that in the ordinary nonprobabilistic Boolean logic strong
disjunction of two propositions in contrast with usual dijunction is true,
in the case when only one from propositions $A,B$ is true and the other is
false). Now let us reduce the expression Eq. (\ref{b4m})\ to more
appropriate form. With this end in view we will consider the representative
matrices of $A$ and $B$ as density matrices of relevant two level quantum
system and write down them in the standard Bloch notation. For example
proposition $A$ \ takes the form: $\rho \left( A\right) =\frac{1}{2}%
\begin{pmatrix}
1+P_{z} & P_{x}-iP_{y} \\ 
P_{x}+iP_{y} & 1-P_{z}%
\end{pmatrix}%
$ with corresponding Bloch vector $P$. Comparing this notation with
expression Eq. (\ref{b2}) we obtain for components of vector $P$ the
relations: $P_{x}=0,\alpha =-\frac{P_{y}}{2}$ and $\ p=\frac{1}{2}(1+P_{z})$%
. We see that the set of GP can be put into one to one correspondence with
the set of various mixed states of two level quantum systems polarized in
plane $Y-Z$. It is convinient to introduce the complex vector $%
P=P_{z}-iP_{y} $ ( we will continue to call it the Bloch vector of the
state). Now one can easy verify directly that if proposition $A$ has the
Bloch vector $P$, than the negation $\overline{A}$ has the Bloch vector $(-P)
$ and proposition $(A\Delta B)$ in this notation can be represented in the
form $\rho ($ $A\Delta B)\equiv \frac{1}{2}%
\begin{pmatrix}
1+R_{z} & -iR_{y} \\ 
iR_{y} & 1-R_{z}%
\end{pmatrix}%
$ with complex vector $R=-PQ$. The components of the Bloch vector $R$ of the
proposition $(A\Delta B)$ can be written as : $R_{z}=P_{y}Q_{y}-P_{z}Q_{z}$
and $R_{y}=-\left( P_{y}Q_{z}+P_{z}Q_{y}\right) .$It is worth noting here
two simple properties of operation $\Delta $ , which can be easily deduced
from the expression Eq. (\ref{b4m}):%
\begin{eqnarray}
\text{1) }\left( \overline{A\Delta B}\right) &=&\left( \overline{A}\Delta
B\right) =\left( A\Delta \overline{B}\right)  \notag \\
&&\text{and }  \label{b6} \\
\text{2) }\overline{A}\Delta \overline{B} &=&A\Delta B.  \notag
\end{eqnarray}%
It should be noted also, that in contrast with the probabilistic Boolean
logic, in the case of DCL it is possible, aside from mentioned above
discrete \ logical connectives, to define also the additional one parameter
group of continuous logical operations.Indeed,one can make the rotation of
the Bloch vector $P$ of every proposition $A$ in the plane $Y-Z$ \ at an
arbitrary angle $\phi $ that results in to the another proposition $A_{1}$
with corresponding Bloch vector $\ P_{1}$ that has the components:%
\begin{eqnarray}
P_{1y} &=&P_{y}\cos \phi +P_{z}\sin \phi  \notag \\
P_{1z} &=&P_{z}\cos \phi -P_{y}\sin \phi  \label{b7}
\end{eqnarray}%
The expressions Eq. (\ref{b4m}) and Eq. (\ref{b7}) imply that if one rotates
the proposition $A$ at an angle $\phi _{1}$ and the proposition $B$ at an
angle $\phi _{2}$ the proposition $(A\Delta B)$ \ is rotated at an angle $%
\phi =(\phi _{1}+\phi _{2})$. Now let us compare the "logical power" of DCL
with the "power" of ordinary probabilistic Boolean logic. At first sight it
may seem that in certain respects DCL is more poor logical theory because it
posseses only four discrete connectives while the standard Boolean logic has
16 similar connectives. However such conclusion would be hasty. To
demonstrate the great opportunities of DCL we are going first of all to
prove that every CP(which is represented by purely diagonal matrix) can be
obtained from two identical GP by the single discrete operation $\Delta $
(strong disjunction). This fact is in contrast with ordinary Boolean logic
where the proposition $(A\Delta A)$ is always false and with probabilistic
Boolean logic where the plausible proposition $(A\Delta A)$ has the
following form: $(A\Delta A)=%
\begin{pmatrix}
2p-2p^{2} & 0 \\ 
0 & 1-2p+2p^{2}%
\end{pmatrix}%
$, and hence its plausibility is always less than one half for any $\ $%
proposition $A$. But if we turn to the case of DCL and take as starting the
proposition $A$ $=%
\begin{pmatrix}
\frac{1}{2} & i\alpha \\ 
-i\alpha & \frac{1}{2}%
\end{pmatrix}%
$ we, using Eq. (\ref{b4m}) result in that strong disjunction $(A\Delta A)$
has the form: $(A\Delta A)=%
\begin{pmatrix}
\frac{1}{2}+2\alpha ^{2} & 0 \\ 
0 & \frac{1}{2}-2\alpha ^{2}%
\end{pmatrix}%
$ and hence any plausible proposition, whose plausibility is more than $%
\frac{1}{2}$, can be obtained in this way by an appropriate choice of $%
\alpha $ (recall that parameter $\alpha $ takes its values in the interval $%
\left( -\frac{1}{2}\leq \alpha \leq \frac{1}{2}\right) $). Note that an
arbitrary CP can be represented in the form of strong disjunction in two
ways (since the relation $A\Delta A=\overline{A}\Delta \overline{A}$ is
identically holds as we have stated earlier). The proven result gives one
the good reason to presume that DCL may precede (as the possible prime
substructure) to the ordinary Boolean logic, which is, however, recorded
better by our consciousness. It should be noted also that not only classical
but arbitrary GP $A$ can be represented as strong disjunction of two
identical propositions, namely: $A=\left( A_{1}\Delta A_{1}\right) $ for
some proposition $A_{1}.$The determination of $A_{1\text{ }}$can be
considered, in figurative sense, as taking the logical squire root from
proposition $A$, since if proposition $A$ has Bloch vector $R$ and
proposition $A_{1}$ has Bloch vector $P$ then the relation $R=-P^{2\text{ }}$
holds as it follows from the foregoing text. One can explicitly write down
this relation in components and solve it but we do not dwell on this point.
On the other hand the presence of additional one parametric group of
continuous operations in the DCL in our opinion greatly expends its
opportunities as the tool for various logical and information processing.
This fundamental issue certainly deserves of special \ attention and
research but in this paper we restrict ourselves only to the remark that for
example, both the concept of thought rotations and the theory of mental
imagery which were introduced into cognitive psychology mainly by R. Shepard 
$\ $and S.Kosslin (see e.g. \cite{5s}, \cite{6s}) can be naturally
interpreted in the language of similar logical operations.In the remainder
of the paper we are going to demonstrate how formal constructions of DCL,
described above, can be implemented in relevant quantum systems by the
modern quantum engeneering tools. Note that our exposition of this problem
will be to a large extent sketchy. To get acquanted with the opportunities
of this technique at lengh we recommend the reader turn to the book \cite{7s}%
. So, let us consider several concrete tasks that everyone should be able to
perform for simulating main DCL constructions. Obviously first of all one
need to create a sufficient reserve of physical realizations of GP. To this
end in view it is naturally to use the states of open two- level quantum
system whose polarization vector is situated in the $Y-Z$ plane. A natural
way to create such states is to organize the interaction of the open system
with its environment suchwise that a component $P_{x}$ of initial mixed
state rapidly decayed with time. Here we specify only one simple way to
achieve this goal. Let us use for the description of evolution of open
quantum system the well-known Lindblad master equation that in general case
has the following form:%
\begin{equation}
\frac{d\rho }{dt}=-\frac{i}{\hbar }\left[ H,\rho \right] +\sum%
\limits_{j=1}^{N}\ \left[ R_{j}\rho ,\text{ }R_{j}^{+}\right] +h.c,
\label{b8}
\end{equation}%
(where $H$ is some hermitian operator and operators $R_{j}$, $R_{j}^{+}$ are
a set, generally speaking, nonhermitian operators). Taking jointly these
operators describe both internal dynamics of the system in question and its
interaction with environment. In the case of two- level quantum system, that
we are only interested in, it is convinient to use the Bloch representation
for it density matrix, namely: $\rho =\frac{1+\overrightarrow{P}%
\overrightarrow{\sigma }}{2}$, where $\overrightarrow{P}$ is polarization
vector of the state, and $\overrightarrow{\sigma }=\left\{ \sigma
_{i}\right\} $ ($($ $i=1,2,3)-$ are standard Pauli matrices). Taking into
account that any $2\times 2$ hermitian matrix can be decomposed in Pauli
matrices one can write down all operators entering in Eq. (\ref{b8}) in the
following form: $H=2\overrightarrow{h}\overrightarrow{\sigma }$, and $R_{j}=%
\overrightarrow{A_{j}}\overrightarrow{\sigma }+i\overrightarrow{B_{j}}%
\overrightarrow{\sigma }$. The set of vectors:$\overrightarrow{h}$, $%
\overrightarrow{A_{j}}$, $\overrightarrow{B_{j}}$ are completely
characterizes the dynamics of two- level open system within the Lindblad
equation approach.Based on Eq. (\ref{b8}) after elementary algebra one can
obtain the required evolution equation for the polarization vector $%
\overrightarrow{P}$ that reads as:%
\begin{widetext}
\begin{equation}
\frac{d\overrightarrow{P}}{dt}=\left( \overrightarrow{h}\times 
\overrightarrow{P}\right) +\sum\limits_{j=1}^{N}2\left( \overrightarrow{A_{j}%
}\times \overrightarrow{B_{j}}\right) -\overrightarrow{A_{j}}\times \left( 
\overrightarrow{P}\times \overrightarrow{A_{j}}\right) -\overrightarrow{B_{j}%
}\times \left( \overrightarrow{P}\times \overrightarrow{B_{j}}\right) .
\label{b9}
\end{equation}
\end{widetext}Let us consider the simplest situation when $N=1$ and $%
\overrightarrow{h}=0$ and choose as $R$ \ the hermitian operator : $R=%
\overrightarrow{A}\overrightarrow{\sigma }$

In this case the Eq. (\ref{b9}) for the polarization vector $P$ takes the
simple form:%
\begin{equation}
\frac{d\overrightarrow{P}}{dt}=-\left( \overrightarrow{A}\times \left( 
\overrightarrow{P}\times \overrightarrow{A}\right) \right) .  \label{b10}
\end{equation}%
It is clear that evolution of the system according to Eq. (\ref{b10})
physically corresponds to the continuous measurement of the observable $%
O\equiv \overrightarrow{A}\overrightarrow{\sigma }$ in an arbitrary
two-level system.Therefore if one will choose the observable $O$ in the
form: $O_{i}=a_{i}\sigma _{y}+b_{i}\sigma _{z}$ (where $a_{i}$ and $b_{i}$
some numerical coefficients ) the terminal state of given open system will
simulate a certain proposition from DCL (since its polarization vector will
be situated in the $Y-Z$ plane). Varying the constants $a_{i\text{ }}$ and $%
b_{i}$ by appropriate way, one is able to create initial reserve of relevant
mixed states which are the physical realizations of required generalized
propositions. The next necessary stage towards the realization of main
constructions of the DCL is a simulation of various logical operations with
these states, that are considered now as already present. Among these
operations there are certain unitary operations such as negation and all
logical rotations. They will not be considered in this paper, since it is
well known that any unitary operator can be realized by the relevant quantum
circuit consisting of some universal gates (see e.g. \cite{8s}).Therefore we
will discuss here only the method of implementation relating to the single
operation,namely, the strong disjunction which ,clearly, can not be reduced
to any unitary operator. Remind that, as it was shown above, the strong
disjunction is determined by the following not quadratic $2\times 4$ matrix: 
$G_{\Delta }=%
\begin{pmatrix}
0 & 1 & 1 & 0 \\ 
1 & 0 & 0 & 1%
\end{pmatrix}%
$ that acts on the tensor product $A\otimes B$ of two propositions $A=%
\begin{pmatrix}
p & i\alpha  \\ 
-i\alpha  & 1-p%
\end{pmatrix}%
$ and $B=%
\begin{pmatrix}
q & i\beta  \\ 
-i\beta  & 1-q%
\end{pmatrix}%
$ which in explicit form has the form:%
\begin{widetext}
\begin{equation}
A\otimes B=%
\begin{pmatrix}
pq & i\beta p & i\alpha q & -\alpha \beta  \\ 
-i\beta p & p\left( 1-q\right)  & \alpha \beta  & i\alpha \left( 1-q\right) 
\\ 
-i\alpha q & \alpha \beta  & q\left( 1-p\right)  & i\beta \left( 1-p\right) 
\\ 
-\alpha \beta  & -i\alpha \left( 1-q\right)  & -i\beta \left( 1-p\right)  & 
\left( 1-p\right) \left( 1-q\right) 
\end{pmatrix}%
,  \label{b11}
\end{equation}
\end{widetext}according to the rule: $\left( A\Delta B\right) =G_{\Delta }%
\left[ A\otimes B\right] G_{\Delta }^{T}$. Since, it is difficult to realize
by physical means operations with nonquadratic matrices, one can use instead
of $2\times 4$ matrix G$_{\Delta }$ the following $4\times 4$ matrix: $G_{4}=%
\begin{pmatrix}
0 & 0 & 0 & 0 \\ 
0 & 1 & 1 & 0 \\ 
0 & 0 & 0 & 0 \\ 
1 & 0 & 0 & 1%
\end{pmatrix}%
$. It easy to verify directly that if we create the state of composite
quantum system with density matrix :$\rho _{4}=G_{4}\left( A\otimes B\right)
G_{4}^{T},$ that has the following explicit form:%
\begin{equation}
\rho _{4}=%
\begin{pmatrix}
0 & 0 & 0 & 0 \\ 
0 & \rho _{22} & 0 & \rho _{24} \\ 
0 & 0 & 0 & 0 \\ 
0 & \rho _{42} & 0 & \rho _{44}%
\end{pmatrix}%
,  \label{b12}
\end{equation}%
(where $\rho _{22}=p+q-2pq+2\alpha \beta $, $\rho _{24}=\rho _{42}^{\ast
}=i\alpha \left( 1-2q\right) +i\beta \left( 1-2p\right) $, $\rho
_{44}=1-\rho _{22\text{ }}$) and then after that take the projection of $%
\rho _{4}$ in the first subsystem we obtain exactly the desired matrix $\rho
\left( A\Delta B\right) $ from Eq. (\ref{b4m}). However, there is still a
problem: how to realize in physical system the nonunitary operator $G_{4}$?
Clearly, we can realize this nonunitary operation in the composite system
only in the case when our system of interest would interact with environment
in a special chosen manner. We point out here one concrete method to
accomplish the required task, but we are aware that this method is neither
unique and probably nor the simplest. So, let us assume that the evolution
of relevant quantum system with initial density matrix Eq. (\ref{b11}) is
determined by the special system of linear differential equations for its
matrix elements.It is convinient to decompose these equations into two
groups:the first group for the diagonal elements and the second group for
nondiagonal matrix elements only. The equations of the first group have the
following form:%
\begin{eqnarray}
\frac{d\rho _{11}}{dt} &=&-\frac{\rho _{11}}{\tau _{1}},\frac{d\rho _{33}}{dt%
}=-\frac{\rho _{33}}{\tau _{3}},  \notag \\
\frac{d\rho _{22}}{dt} &=&\frac{\rho _{33}}{\tau _{3}}-\frac{\rho _{14}}{%
\tau _{14}}+\frac{\rho _{23}}{\tau _{23}},  \label{b13m} \\
\frac{d\rho _{44}}{dt} &=&\frac{\rho _{11}}{\tau _{1}}+\frac{\rho _{14}}{%
\tau _{14}}-\frac{\rho _{23}}{\tau _{23}},  \notag
\end{eqnarray}%
and the equations of the second group that read as:%
\begin{eqnarray}
\frac{d\rho _{12}}{dt} &=&-\frac{\rho _{12}}{\tau _{12}},\frac{d\rho _{13}}{%
dt}=-\frac{\rho _{13}}{\tau _{13}},  \notag \\
\frac{d\rho _{14}}{dt} &=&-\frac{\rho _{14}}{\tau _{14}},\frac{d\rho _{23}}{%
dt}=-\frac{\rho _{23}}{\tau _{23}}  \label{b14} \\
\frac{d\rho _{24}}{dt} &=&-\frac{\rho _{12}}{\tau _{12}}-\frac{\rho _{13}}{%
\tau _{13}}+\frac{\rho _{34}}{\tau _{34}},  \notag \\
\frac{d\rho _{34}}{dt} &=&-\frac{\rho _{34}}{\tau _{34}},  \notag
\end{eqnarray}%
(in equations Eq. (\ref{b13m}) and Eq. (\ref{b14}) we use the notation $\tau
_{i}$ $,\tau _{ik}$ for the relaxation times of corresponding states and
transitions ). We do not write down here the similar equations for the
remaining matrix elements $\rho _{ik}$ since we assume that their evolution
is determined uniquely by the principle of detailed balance. The elementary
analysis of Eq. (\ref{b12}), and Eq. (\ref{b13m}) shows that when t tends to
infinity (that is for the stationary density matrix) diagonal matrix
elements $\rho _{11}$, $\rho _{33}$ and also nondiagonal matrix elements:$%
\rho _{12}$, $\rho _{13}$, $\rho _{14}$, $\rho _{23}$, $\rho _{34\text{ }}$%
vanish. Note that the system Eq. (\ref{b12}), Eq. (\ref{b13m}) have two
integrals of motion namely $I_{1}=\rho _{22}+\rho _{33}+\rho _{23}-\rho _{14}
$ and $I_{2}=\rho _{24}+\rho _{34}-\rho _{12}-\rho _{13}$. Taking into
account that in initial state $I_{1}=p+q-2pq+2\alpha \beta $ and $%
I_{2}=i\alpha \left( 1-2q\right) $ +$i\beta \left( 1-2p\right) $ one can
conclude that for terminal state $\rho _{22}=I_{1}=p+q-2pq+2\alpha \beta $
and $\rho _{24}=i\alpha \left( 1-2q\right) +i\beta \left( 1-2p\right) .$
QED. Thus we prove that it is possible to organize the interaction of open
composite quantum system with its environment suchwise that its initial
state $A\otimes B$ eventually transforms in required state namely $\rho
\left( A\Delta B\right) $ in one of its subsystems.It should be noted
although we describe the relevant evolution of the system directly in the
language of differential equations the same result can be reached also by
the appropriate Lindblad equation since the open quantum system of interest
is undoubtedly the quantum Markov system.

In conclusion we want emphasize once more that the main subject of our study
in present paper is , using the semiotic language, only the syntax of the
DCL. The other equally important issues relating to semantics of this
theory, that is an interpretation of all its basic constructions remained
out of our scope. Also we are not touched the possible concrete applications
of the DCL in physics although , for example, the study of the quantum logic
problems from this point of view suggests itself. All mentioned issues we
hope to study and discuss in detail in our further publications.

\end{document}